\documentclass[a4paper,12pt]{article}
\usepackage{amsmath}
\usepackage{amssymb}
\usepackage{amsfonts}
\usepackage{fleqn} 
\usepackage{mathrsfs}

\def\beq{\begin{equation}}
\def\eeq{\end{equation}}
\def\bea{\begin{eqnarray}}
\def\eea{\end{eqnarray}}

\def\nuL{\nu}
\def\EL{\mathrm{e}}

\def\<{\left\langle}
\def\>{\right\rangle}

\begin{document}

\bibliographystyle{OurBibTeX}

\begin{titlepage}

 \vspace*{-15mm}
\begin{flushright}
SHEP/0525\\
hep-ph/0508044
\end{flushright}
\vspace*{5mm}

\begin{center}
{
\sffamily
\Large
Charged Lepton Corrections to Neutrino Mixing Angles\\[2mm]
and CP Phases Revisited
}
\\[12mm]
S.~Antusch\footnote{E-mail: \texttt{santusch@hep.phys.soton.ac.uk}},
S.~F.~King\footnote{E-mail: \texttt{sfk@hep.phys.soton.ac.uk}}
\\[1mm]
{\small\it
School of Physics and Astronomy,
University of Southampton,\\
Southampton, SO17 1BJ, U.K.
}
\end{center}
\vspace*{1.00cm}

\begin{abstract}

\noindent 
We re-analyze charged lepton corrections to neutrino mixing angles and
CP phases, carefully including CP phases from the charged lepton
sector. We present simple analytical formulae for including the
charged lepton corrections and derive compact new results for small
neutrino and charged lepton mixings $\theta^{\nuL}_{13}$ and
$\theta^{\EL}_{13}$. We find a generic relation $ \theta_{12} +
\frac{1}{\sqrt{2}} \theta^{\EL}_{12} \cos(\delta - \pi) \approx
\theta^{\nuL}_{12} $, which relates the prediction from the neutrino
sector $\theta^{\nuL}_{12}$ to the charged lepton mixing
$\theta^{\EL}_{12}$ and to the MNS neutrino oscillation phase
$\delta$. We apply our formula to the examples of bimaximal or
tri-bimaximal neutrino mixing. One implication is that the so-called
quark-lepton complementarity relation $\theta_{12} + \theta_\mathrm{C}
= 45^\circ$ can only hold for $\delta = \pi$ and it gets modified in
the presence of leptonic CP violation. On the other hand, the lepton
mixing $\theta_{13}$ generated from the charged lepton correction
$\theta^{\EL}_{12}$ is independent of CP phases and given by
$\theta_{13} = \frac{1}{\sqrt{2}} \theta^{\EL}_{12}$.  Combining these
results leads to a model-independent sum rule: $\theta_{12} +
\theta_{13}\cos(\delta - \pi) \approx \theta^{\nuL}_{12} $ where
$\theta^{\nuL}_{12} = (35.26^\circ)\: 45^\circ$ in the case of
(tri-)bimaximal neutrino mixing, for example.

\end{abstract}

\end{titlepage}
\newpage
\setcounter{footnote}{0}

\section{Introduction}
Recently, there has been some interest in relations between the lepton
mixing angle $\theta_{12}$ and the Cabibbo angle $\theta_\mathrm{C}$
\cite{QLCliterature}, often referred to as quark-lepton
complementarity (QLC).  It was initiated from the observation that
current best fit values (see e.g.\ \cite{Maltoni:2004ei}) are
compatible with an intriguing relation $\theta_{12}
+\theta_\mathrm{C}=45^\circ$.
One conclusion was that if a relation between the Cabibbo angle and
$\theta_{12}$ were found, this could point towards quark-lepton
unification. In addition, further relations between the lepton mixings
$\theta_{13}$ and $\theta_{23}$ and quark mixings have been proposed.
The general idea is that specific predictions from the neutrino
sector, i.e.\ for the solar angle, get modified by the charged lepton
mixings \cite{ChargedLeptonCorrections} such as $\theta^{\EL}_{12}$,
and that these corrections from the charged lepton sector can be be
related to quark mixings in quark-lepton unified theories.
While the above relation, where $\theta_{12}$ and $\theta_\mathrm{C}$
add up to $45^\circ$, was based on bimaximal neutrino mixing, another
interesting complementarity emerges if the neutrino sector predicts
tri-bimaximal mixing \cite{tribi}, which can naturally be realized
with non-Abelian flavour symmetry SO(3) \cite{King:2005bj} or SU(3)
\cite{deMedeirosVarzielas:2005ax}.

Motivated by this recent interest in charged lepton corrections to
neutrino mixings, we re-analyze this issue in this note. We shall
first present simple formulae for including the charged lepton
corrections, which allow to discuss the effects analytically. Since
the two types of complementarity scenarios involve small 1-3 mixing
$\theta^{\nuL}_{13}$ from the neutrino sector, we then mainly focus on
this case and find some interesting results: For small neutrino and
charged lepton mixings $\theta^{\nuL}_{13}$ and $\theta^{\EL}_{13}$,
we find for instance the new relation
\begin{eqnarray}\label{Eq:NewRelationForT12}
\theta_{12} + \frac{1}{\sqrt{2}} \theta^{e}_{12} \cos(\delta - \pi)
\approx 
 \theta^{\nu}_{12}\; , 
\end{eqnarray}
which connects the prediction $\theta^{\nuL}_{12}$ for the solar
mixing from the neutrino sector to the charged lepton mixing
$\theta^{\EL}_{12}$ and to the MNS CP phase relevant for neutrino
oscillations $\delta$. Compared to previous works which consider
charged lepton contributions \cite{ChargedLeptonCorrections}, the
general relation of Eq.~(\ref{Eq:NewRelationForT12}) holds
model-independently as long as the neutrino and charged lepton mixings
$\theta^{\nuL}_{13}$ and $\theta^{\EL}_{13}$ are small.  It is
surprisingly compact and shows that under the above conditions,
measuring the MNS CP phase $\delta$ is essential for testing any model
predictions for the neutrino mixing $\theta^{\nuL}_{12}$ modified by
charged lepton corrections.

The compact formula in Eq.~(\ref{Eq:NewRelationForT12}) may be
used as a basis for studying quark-lepton complementarity, if
the charged lepton mixing angle $\theta^{e}_{12}$ is related to the
Cabibbo angle $\theta_C$.  The exact type of complementarity depends
on the prediction for the neutrino mixing angle $\theta^{\nuL}_{12}$,
specifically either $\theta^{\nu}_{12}\approx 45^\circ$ in the case of
bi-maximal complementarity, or $\theta^{\nu}_{12}\approx 35.26^\circ$
in the case of tri-bimaximal complementarity. In both cases,
Eq.~(\ref{Eq:NewRelationForT12}) shows that the presence of the
leptonic CP phase $\delta$ plays a crucial role in any complementarity
relation. Indeed bimaximal complementarity in its simple form is seen
to be disfavored in the presence of leptonic CP violation (i.e.\
$\delta \not= \pi$). Tri-bimaximal complementarity gives testable
predictions, which however require a measurement of $\delta$ in
addition to a more precise measurement of $\theta_{12}$. In the
context of tri-bimaximal complementarity \cite{King:2005bj}, the
relation of Eq.~(\ref{Eq:NewRelationForT12}) with the specific
prediction for $\theta^{\nu}_{12}\approx 35.26^\circ$ has been found
in a specific model.  Discussions of charged lepton contributions to
neutrino mixings in other specific scenarios, such as for instance
bi-maximal neutrino mixing, can be found in Refs.\
\cite{ChargedLeptonCorrections}.  However, as we will show, the
general relation of Eq.~(\ref{Eq:NewRelationForT12}) holds
model-independently, as long as $\theta^{\nuL}_{13}$ and
$\theta^{\EL}_{13}$ are small.

For small $\theta^{\nuL}_{13}$ and $\theta^{\EL}_{13}$, the total lepton 
mixing $\theta_{13}$ is induced from the charged lepton correction 
$\theta^{\EL}_{12}$, which leads to the relation
\begin{eqnarray}
\theta_{13} = \frac{1}{\sqrt{2}} \theta^{e}_{12}\;,
\end{eqnarray}
independent of CP phases. This means that if the charged lepton mixing
$\theta^{\EL}_{12}$ is related to the Cabibbo angle $\theta_{\mathrm{C}}$ 
in any form, this
would show up more directly in $\theta_{13}$ than in the solar angle 
$\theta_{12}$. 

Combining these results leads to a model-independent sum rule:
\beq
\theta_{12} + \theta_{13}\cos(\delta - \pi) \approx
\theta^{\nu}_{12} 
\label{sumrule}
\eeq
where $\theta^{\nu}_{12}=45^\circ$
in the case of bimaximal neutrino mixing, or 
$\theta^{\nu}_{12}=35.26^\circ$
in the case of tri-bimaximal neutrino mixing, for example. 
It is worth emphasizing that under the generic assumption of small $\theta^{\nuL}_{13}$ and $\theta^{\EL}_{13}$ the combined measurement of the lepton mixings $\theta_{12}$, $\theta_{13}$ and of the MNS CP phase $\delta$ in future precision experiments on neutrino oscillations has the potential to reveal if there are any symmetries determining the neutrino mixing $\theta^{\nu}_{12}$.

In the most general case, if we relax the condition of small $\theta^{\nuL}_{13}$ and 
$\theta^{\EL}_{13}$, charged lepton CP phases still modify the charged
lepton corrections to the solar mixing angle, however the relevant CP phase is
then not related to the low energy CP phase $\delta$ observable (in principle)
in future neutrino oscillation experiments. Then the situation is similar to the
charged lepton correction to $\theta_{23}$: Since it depends on charged lepton
CP phases which are not related to $\delta$ and just marginally contribute to
one of the Majorana CP phases, we conclude that it is not realistic to expect 
any generic complementarity relation for $\theta_{23}$. The maximal charged
lepton correction to $\theta_{23}$ is $|\Delta \theta_{23}| \lesssim 
\theta^{\EL}_{23}$, which is nevertheless interesting with respect to future
precision neutrino experiments.

\section{Preliminaries on the Mixing Formalism}\label{conventions}
Before we discuss charged lepton corrections, it is necessary to 
specify the definition of lepton mixings and our conventions for the charged 
lepton and neutrino mass matrices:  
The Dirac mass matrices of the charged leptons
is given by 
\begin{eqnarray}
m^{\mathrm{e}}_\mathrm{LR}=Y^\mathrm{e}_\mathrm{LR}v_\mathrm{d}
\end{eqnarray}
where
$v_\mathrm{d} = \< h^0_\mathrm{d}\>$ 
and the Lagrangian is of the form 
\begin{eqnarray}
{\cal L}=-Y^\mathrm{e}_\mathrm{LR} \bar{e}_L  h e_R + \text{H.c.}
\end{eqnarray}
We will focus on three light Majorana neutrinos in the following, with the 
neutrino mass
being defined by the Lagrangian
\begin{eqnarray}
{\cal L}=-\tfrac{1}{2}\bar{\nu}_L m^{\nu}_{LL} \nu_\mathrm{L}^c + \text{H.c.}
\end{eqnarray} 
The change from flavour basis to mass
eigenbasis can be performed with the unitary diagonalization matrices
$V_{\mathrm{e}_\mathrm{L}},V_{\mathrm{e}_\mathrm{R}}$ and
$V_{\nu_\mathrm{L}}$ by
\begin{eqnarray}\label{eq:DiagMe}
V_{\mathrm{e}_\mathrm{L}} \, m^{\mathrm{e}}_\mathrm{LR} \,
V^\dagger_{\mathrm{e}_\mathrm{R}} =
\left(\begin{array}{ccc}
\!m_e&0&0\!\\
\!0&m_\mu&0\!\\
\!0&0&m_\tau\!
\end{array}
\right)\! , \quad
V_{\nu_\mathrm{L}} \,m^\nu_{\mathrm{LL}}\,V^T_{\nu_\mathrm{L}} =
\left(\begin{array}{ccc}
\!m_1&0&0\!\\
\!0&m_2&0\!\\
\!0&0&m_3\!
\end{array}
\right)\! .
\end{eqnarray}
The MNS matrix, the mixing matrix in the lepton sector, is then given by
\begin{eqnarray}\label{Eq:MNS_Definition}
U_{\mathrm{MNS}} = V_{e_\mathrm{L}} V^\dagger_{\nu_\mathrm{L}}\; .
\end{eqnarray}
After eliminating so-called unphysical phases as usual, by 
charged lepton phase rotations, the MNS matrix can be
parameterized as 
\begin{eqnarray}
U_{\mathrm{MNS}} = R_{23} U_{13} R_{12} P_0
\end{eqnarray}
using the matrices
 $R_{23}, U_{13}, R_{12}$ and $P_0$ defined by 
\begin{align}\label{eq:R23U13R12P0}
R_{12}:=
\left(\begin{array}{ccc}
  c_{12} & s_{12} & 0\\
  -s_{12}&c_{12} & 0\\
  0&0&1\end{array}\right)
  , \:&
\quad U_{13}:=\left(\begin{array}{ccc}
   c_{13} & 0 & s_{13}e^{-i\delta}\\
  0&1& 0\\
  - s_{13}e^{i\delta}&0&c_{13}\end{array}\right)  ,  \nonumber \\
R_{23}:=\left(\begin{array}{ccc}
 1 & 0 & 0\\
0&c_{23} & s_{23}\\
0&-s_{23}&c_{23}
 \end{array}\right)
  , \:&
 \quad P_0:=
 \begin{pmatrix}
 e^{i\beta_1}&0&0\\0&e^{i\beta_2}&0\\0&0&1
  \end{pmatrix}
\end{align}
and where $s_{ij}$ and $c_{ij}$ stand for $\sin (\theta_{ij})$ and $\cos
(\theta_{ij})$, respectively.
The matrix $P_0$ contains the Majorana
phases $\beta_1$ and $\beta_2$.
 $\delta$ is the Dirac CP phase relevant for neutrino oscillations.

Another useful parameterization, in particular for including charged lepton
corrections, is \cite{King:2002nf}
\begin{eqnarray}
U_{\mathrm{MNS}} = U_{23} U_{13} U_{12}\; ,
\end{eqnarray}
with matrices $U_{23}, U_{13}, U_{12}$ being defined as
\begin{align}\label{eq:U23U13U12}
  U_{12}:=
\left(\begin{array}{ccc}
  c_{12} & s_{12}e^{-i\delta_{12}} & 0\\
  -s_{12}e^{i\delta_{12}}&c_{12} & 0\\
  0&0&1\end{array}\right)
  , \:&
\quad U_{13}:=\left(\begin{array}{ccc}
   c_{13} & 0 & s_{13}e^{-i\delta_{13}}\\
  0&1& 0\\
  - s_{13}e^{i\delta_{13}}&0&c_{13}\end{array}\right)  ,  \nonumber \\
U_{23}:=
\left(\begin{array}{ccc}
 1 & 0 & 0\\
0&c_{23} & s_{23}e^{-i\delta_{23}}\\
0&-s_{23}e^{i\delta_{23}}&c_{23}
 \end{array}\right). &
\end{align}
One can easily switch between these two conventions 
using the identities \cite{King:2002nf}
\begin{subequations}\label{Eq:Dict}\begin{eqnarray}
\label{Eq:Dict23}\delta_{23} &=& \beta_2 \\
\label{Eq:Dict13}\delta_{13} &=& \delta + \beta_1 \\
\label{Eq:Dict12}\delta_{12} &=& \beta_1 - \beta_2 \; .
\end{eqnarray}\end{subequations}
and the fact that $\theta_{ij}$ remains the same in both notations. 
We will use the latter convention in the following and introduce the more common phase
convention $\delta,\beta_1,\beta_2$ if appropriate, in particular for making the
connection to the Dirac CP phase $\delta$ observable in neutrino oscillations.

\section{Simple Formulae for Including Charged Lepton Corrections}
We will now consider the situation that bi-large neutrino mixing stems mainly from
the neutrino sector, and that the mixing angles induced by the charged leptons
can be considered as corrections. In this approximation, we will 
derive formulae which allow to include corrections to neutrino mixing angles 
and CP phases conveniently. We will see that special care has to be taken when
dealing with complex phases from the charged lepton sector.  

Parameterizing the neutrino and charged lepton diagonalization matrices in 
$V^\dagger_{e_\mathrm{L}}$ and $V^\dagger_{\nu_\mathrm{L}}$ an analogous way to
Eq.~(\ref{eq:U23U13U12}),  
we can write $U_{\mathrm{MNS}}$ as \cite{King:2002nf}
\begin{eqnarray}
U_{\mathrm{MNS}} = U^{\mathrm{e}_\mathrm{L}\dagger}_{12} U^{\mathrm{e}_\mathrm{L}\dagger}_{13} 
U^{\mathrm{e}_\mathrm{L}\dagger}_{23}  U^{\nu_\mathrm{L}}_{23}
U^{\nu_\mathrm{L}}_{13} U^{\nu_\mathrm{L}}_{12} \; .
\end{eqnarray}  
The additional unphysical phases have been shifted to the left and then
absorbed, as usual, by charged lepton phase rotations. 
The procedure for extracting the charged lepton and neutrino 
angles and phases is given in great detail 
in the Appendix of Ref.~\cite{King:2002nf}.  

In this parameterization, the MNS matrix can be conveniently expanded
in terms of neutrino and charged lepton mixing angles and phases
to leading order in small quantities, i.e.\ in the 
charged lepton mixing angles and in $\theta^{\nuL}_{13}$:
\footnote{These results differ somewhat from those in
  \cite{King:2002nf}, in particular
the sign of the last term in Eq.~(\ref{chlep12}) has been corrected.}
\begin{subequations}\bea 
\label{Eq:23} s_{23}e^{-i\delta_{23}}
& \approx &
s_{23}^{\nuL}e^{-i\delta_{23}^{\nuL}}
-\theta_{23}^{\EL}
c_{23}^{\nuL}e^{-i\delta_{23}^{\EL}}
\label{chlep23}
\\
\label{Eq:13} \theta_{13}e^{-i\delta_{13}}
& \approx &
\theta_{13}^{\nuL}e^{-i\delta_{13}^{\nuL}}
-\theta_{13}^{\EL}c_{23}^{\nuL}e^{-i\delta_{13}^{\EL}}
-\theta_{12}^{\EL}s_{23}^{\nuL}e^{i(-\delta_{23}^{\nuL}-\delta_{12}^{\EL})}
\label{chlep13}
\\
\label{Eq:12} s_{12}e^{-i\delta_{12}}
& \approx &
s_{12}^{\nuL}e^{-i\delta_{12}^{\nuL}}
+\theta_{13}^{\EL}
c_{12}^{\nuL}s_{23}^{\nuL}e^{i(\delta_{23}^{\nuL}-\delta_{13}^{\EL})}
- \theta_{12}^{\EL} c_{23}^{\nuL}c_{12}^{\nuL}e^{-i\delta_{12}^{\EL}}
\label{chlep12}
\eea\end{subequations}
Using Eq.~(\ref{Eq:Dict}) it is simple to express the phases on the left-hand
side in terms of the phases $\beta_1,\beta_2$ and $\delta$, if desired. 
Before we turn to
applications, let us remark that since we have assumed that the two large lepton mixing angles
$\theta_{23}$ and $\theta_{12}$ stem mainly from the neutrino sector, the phases
$\delta_{23},\delta_{12}$ and thus also $\beta_1,\beta_2$ are mainly determined 
 from the neutrino sector, with only small corrections from the charged lepton
 sector.   
 
With this respect, $\theta_{13}$ and the Dirac CP phase $\delta$ are 
very different: Since we only know that the total lepton mixing $\theta_{13}$ 
is rather small and without making further assumptions, it can stem from the
neutrino mixing $\theta^{\nuL}_{13}$ or it can alternatively be mainly 
induced from the charged lepton mixings $\theta^{\EL}_{13}$ and/or 
$\theta^{\EL}_{12}$. 
In the latter case, the charged lepton corrections also mainly determine the
Dirac CP phase $\delta$, observable in neutrino oscillations.      

Let us finally note that lepton mixing angles are subject to 
renormalization group (RG) running between high energy,
where models typically predict the flavour structure, and low energy, where
experiments are performed. 
A numerical calculation of the RG corrections can be
performed with the software packages REAP/MPT introduced in 
\cite{Antusch:2005gp}.

\subsection{Charged Lepton Corrections with Small $\boldsymbol{\theta^{\nuL}_{13}}$}\label{Sec:SpecialCase}
As discussed in the introduction, one interesting special case is that 
$\theta^{\nuL}_{13}$ as well as 
$\theta^{\EL}_{13}$ are rather small, i.e.\ $\ll \theta^{\EL}_{12}$. 
From Eq.~(\ref{Eq:13}) and using the leading order relations $\delta^{\nuL}_{23}\approx
\beta_2$ and Eq.~(\ref{Eq:Dict13}), we obtain
\begin{eqnarray}
\theta_{13} e^{- i \delta} \approx \theta^{\EL}_{12} s^{\nuL}_{23} 
e^{ - i (\beta_2-\beta_1)-i\pi -i \delta^{\EL}_{12}}\; ,
\end{eqnarray} 
and it follows that
\begin{eqnarray}\label{Eq:delta_1}
\delta \approx \beta_2-\beta_1 + \pi + \delta^{\EL}_{12}\; .
\end{eqnarray}
Let us note that for instance in scenarios with an inverted neutrino mass
hierarchy and a Majorana parity between $m_1$ and $m_2$, we have
$\beta_2-\beta_1 = \pi$ and the Dirac CP phase $\delta$ is simply given by the
charged lepton phase $\delta^{\EL}_{12}$.
In addition, we obtain
\begin{eqnarray}\label{Eq:T13_1}
\theta_{13} =  \theta^{\EL}_{12} s^{\nuL}_{23} \;. 
\end{eqnarray}
The charged lepton 1-2 mixing generates $\theta_{13}$ independent of charged
lepton CP phases.

From Eq.~(\ref{Eq:12}), 
the solar mixing angle $\theta_{12}$ is given by 
\begin{eqnarray}\label{Eq:s12_1}
s_{12} \approx s_{12}^{\nuL}  + \theta^{\EL}_{12} c^{\nuL}_{23} c^{\nuL}_{12} 
\cos(\beta_2 - \beta_1 + \pi + \delta^{\EL}_{12})
\approx s_{12}^{\nuL}  +  \theta^{\EL}_{12} c^{\nuL}_{23} c^{\nuL}_{12} 
\cos( \delta )\; ,
\end{eqnarray}
where we have used the result of Eq.~(\ref{Eq:delta_1}) and 
$\delta_{12} \approx \delta^{\nuL}_{12}$. In terms of mixing angles,
approximating $\theta_{23}^\nuL \approx 45^\circ$, we
obtain:
\begin{eqnarray}\label{Eq:DeltaT12}
\theta_{12} + \frac{1}{\sqrt{2}} \theta^{\EL}_{12} \cos(\delta - \pi)
\approx 
 \theta^{\nuL}_{12} \; .
\end{eqnarray}   
The relation of Eq.~(\ref{Eq:DeltaT12}) 
was first found in a model of tri-bimaximal neutrino mixing based on
SO(3) flavour symmetry \cite{King:2005bj} where $\theta^{\nuL}_{13}=0$ 
and $\theta^{\nuL}_{12}=35.26^\circ$.
We have shown here that such a result holds quite generally,
not just for any model of tri-bimaximal neutrino mixing, but also 
for any model of bimaximal neutrino mixing, or indeed
any model of neutrino mixing in which $\theta^{\nuL}_{13}$ 
and $\theta^{\EL}_{13}$ are small compared to 
$\theta^{\EL}_{12}$.
We emphasize that under these assumptions, the corrections to $\theta_{12}$ from the charged lepton sector depend on the charged lepton 1-2 mixing {\em and} on the
Dirac CP phase $\delta$, a fact which is often overlooked in
studies of complementarity, as we now briefly discuss.

\section{Applications}

\subsection{Consequences for Quark-Lepton Complementarity}
The general idea of quark-lepton complementarity \cite{QLCliterature} is that 
the solar mixing predicted from the neutrino sector is corrected by
charged lepton contributions, which are in turn related to 
quark mixing angles.
For example in the case of bimaximal complementarity,
the starting point is a maximal solar neutrino angle
$\theta^{\nuL}_{12}=45^\circ$, 
and a relation like 
$\theta_{12} + \theta_\mathrm{C} = 45^\circ$, compatible with present best-fit
values, could in principle emerge. For example
in \cite{Antusch:2005ca} it has been shown how a charged lepton mixing
$\theta^{\EL}_{12}=\frac{3}{2}\theta_\mathrm{C}$, which could nearly
realize the QLC relation, can arise in quark-lepton unified theories,
and leads to a prediction for $\theta_{13}\approx \theta_\mathrm{C}$
which also holds in the presence of CP violating phases.
As has been pointed out by many authors, an inverted neutrino mass hierarchy 
with a Majorana parity between $m_1$ and $m_2$ is a good starting point for QLC.
For $\theta^{\nuL}_{12} = 45^\circ$ and
$\theta^{\EL}_{12}=\frac{3}{2}\theta_\mathrm{C}$, under the assumptions of
Sec.~\ref{Sec:SpecialCase} which are in general satisfied in approaches to QLC
via inverted neutrino mass hierarchy, Eq.~(\ref{Eq:DeltaT12}) reads 
\begin{eqnarray}
\theta_{12} + \theta_\mathrm{C}  \cos(\delta - \pi)
\approx 
45^\circ 
\; .
\end{eqnarray}    
In general, we would therefore {\em not} expect a relation 
such as $\theta_{12} + \theta_\mathrm{C} = 45^\circ$, where the Cabibbo angle enters
directly, unless CP is conserved (i.e.\ $\delta=\pi$). 
All quark-lepton complementarity relations are modified for non-zero 
$\delta$ and a measurement of the leptonic Dirac CP phase is required for testing 
Cabibbo-like corrections to neutrino mixing. 
 
An interesting application of our general results 
in the presence of CP violation 
is to tri-bimaximal neutrino mixing, where $\theta^{\nuL}_{13}=0$ 
and $\theta^{\nuL}_{12}=35.26^\circ$. 
For example, for a charged lepton mixing
$\theta^{\EL}_{12}=\theta_\mathrm{C}/3$ corresponding to the
Georgi-Jarlskog relation \cite{Georgi:1979df} which arises in many quark-lepton unified
theories, Eq.~(\ref{Eq:DeltaT12}) leads to a tri-bimaximal complementarity 
relation \cite{King:2005bj} 
\begin{eqnarray}
\theta_{12} + \frac{\theta_\mathrm{C}}{3\sqrt{2}}\cos(\delta - \pi) \approx 
35.26^\circ \; , 
\end{eqnarray}    
which is consistent with current data for a wide range of
CP phases $\delta$.
This also leads to the prediction $\theta_{13}\approx \theta_\mathrm{C}/(3\sqrt{2})$.

\subsection{Consequences for Leptogenesis -- MNS Links}
It is well known that in general, there is no relation between the CP phase
$\delta$ observable in neutrino oscillations and the cosmological CP phase which
appears in the leptogenesis mechanism \cite{Fukugita:1986hr}, where the baryon asymmetry of our
universe arises via out-of-equilibrium decay of the right-handed neutrinos
involved in the see-saw mechanism. 
However, for specific classes of
flavour models with symmetries or specific assumptions such as texture zeros or
sequential dominance conditions, links between these CP violating
phases emerge from predictions for the decay asymmetries $\epsilon_1$
 of the lightest right-handed neutrino. 
Since $\epsilon_1$ depends only on the product $Y_\nu^\dagger Y_\nu$ 
\cite{Covi:1996wh} in the mass basis of the right-handed neutrinos, it is 
not affected by charged lepton
mixings and phases at all. Therefore, Eq.~(\ref{Eq:13}) immediately shows that 
links between the MNS CP phase 
$\delta$ and cosmological CP violation relevant for leptogenesis 
can only hold if 
$\theta^{\EL}_{12}, \theta^{\EL}_{13} \ll \theta^{\nuL}_{13}$, i.e.\ 
if $\theta_{13}$ stems mainly from the neutrino sector. Otherwise, 
there are large contributions to $\delta$ from the charged lepton sector
which are completely decoupled from leptogenesis. If $\theta^{\EL}_{12}$ or 
$\theta^{\EL}_{13}$ are much larger than $\theta^{\nuL}_{13}$, the lepton mixing
$\theta_{13}$ and thus also $\delta$ stem dominantly from charged lepton corrections and any 
leptogenesis-MNS link is lost.

\section{Summary and Conclusions}
In this note, we have revisited charged lepton 
corrections to neutrino mixing and CP phases, carefully including CP 
phases from the charged lepton sector. We have therefore presented simple 
analytical formulae for including the charged lepton corrections. Based on these
formulae, we have derived interesting new results 
for small neutrino and charged lepton mixings $\theta^{\nu}_{13}$ and 
$\theta^{e}_{13}$: For instance, we have found the relation
\begin{eqnarray}
\theta_{12} + \frac{1}{\sqrt{2}} \theta^{e}_{12} \cos(\delta - \pi)
\approx 
 \theta^{\nu}_{12} \nonumber 
\end{eqnarray}
which connects the prediction from the neutrino sector $\theta^{\nu}_{12}$ to 
the charged lepton mixing $\theta^{e}_{12}$ and to the Dirac CP phase
$\delta$. 
We have then applied our formula to the examples of bimaximal or 
tri-bimaximal
neutrino mixing. One implication was that the so-called quark-lepton 
complementarity relation $\theta_{12} + \theta_\mathrm{C} = 45^\circ$ can only hold for
$\delta = \pi$ and it gets modified in the presence of leptonic CP violation.  
We have also found that the lepton mixing $\theta_{13}$ generated from the charged 
lepton correction $\theta^{e}_{12}$ is independent of CP phases and given by 
\begin{eqnarray}
\theta_{13} = \frac{1}{\sqrt{2}} \theta^{e}_{12}\nonumber 
\end{eqnarray}
Combining these results leads to a model-independent sum rule:
\begin{eqnarray}
\theta_{12} + \theta_{13}\cos(\delta - \pi) \approx
\theta^{\nu}_{12} \nonumber
\end{eqnarray}
where $\theta^{\nu}_{12}=45^\circ$
in the case of bimaximal neutrino mixing, or 
$\theta^{\nu}_{12}=35.26^\circ$
in the case of tri-bimaximal neutrino mixing, for example. It is worth emphasizing that under the generic assumption of small $\theta^{\nuL}_{13}$ and $\theta^{\EL}_{13}$ the combined measurement of the lepton mixings $\theta_{12}$, $\theta_{13}$ and of the MNS CP phase $\delta$ in future precision experiments on neutrino oscillations has the potential to reveal if there are any symmetries determining the neutrino mixing $\theta^{\nu}_{12}$.  

In the most general case, if we relax the condition of small $\theta^{\nuL}_{13}$ and 
$\theta^{\EL}_{13}$, charged lepton CP phases still modify the charged
lepton corrections to the solar mixing angle, however the relevant CP phase is
then not related to the low energy CP phase $\delta$ observable (in principle)
in future neutrino oscillation experiments. Then the situation is similar to the
charged lepton correction to $\theta_{23}$. 
The latter depends on charged lepton
CP phases which are not related to $\delta$ and just marginally contribute to
one of the Majorana CP phases, and we conclude that it is not realistic to expect 
any generic complementarity relation for $\theta_{23}$. The maximal charged
lepton correction to $\theta_{23}$ is $|\Delta \theta_{23}| \lesssim 
\theta^{\EL}_{23}$, which is nevertheless interesting with respect to future
precision neutrino experiments.

We have furthermore argued that any link between the MNS CP phase 
$\delta$ and cosmological CP violation relevant for leptogenesis is generically
lost if $\theta^{\nu}_{13}$ is small compared to 
$\theta^{\EL}_{12}$ and/or $\theta^{\EL}_{13}$. Then $\delta$ stems dominantly 
from charged lepton corrections which are completely decoupled from the
 decay asymmetry for leptogenesis.  

\section*{Acknowledgements}
We acknowledge support from the PPARC grant PPA/G/O/2002/00468.

\providecommand{\bysame}{\leavevmode\hbox to3em{\hrulefill}\thinspace}

\end{document}